\documentclass[aps,twocolumn,prl,preprintnumbers,amsmath,amssymb,superscriptaddress]{revtex4-1}

\usepackage{graphicx}
\usepackage{bm}
\usepackage{hyperref}
\usepackage{float}
\usepackage{diagbox}
\usepackage{float}
\usepackage{multibib}
\usepackage{natbib}

 \pdfoutput=1
 
\begin{document} 

\title{Sub-THz momentum drag and violation of Matthiessen's rule in an ultraclean ferromagnetic SrRuO$_3$ metallic thin film}

\author{Youcheng Wang$^1$, G. Boss\'e$^{1,2}$, H. P. Nair$^3$, N. J. Schreiber$^3$, J. P. Ruf$^4$, B. Cheng$^1$, C. Adamo$^3$, D. E. Shai$^4$,
	 Y. Lubashevsky$^1$, D. G. Schlom$^{3,5}$, K. M. Shen$^{4,5}$, and N. P. Armitage$^1$\\
	\medskip
	$^1$ The Institute for Quantum Matter, Department of Physics and Astronomy,
	The Johns Hopkins University, Baltimore, MD 21218 USA\\
	$^2$ Physics Department, University of North Florida, Jacksonville, FL 32224-7699, USA\\
	$^3$ Department of Materials Science and Engineering,
	Cornell University, Ithaca, New York 14853, USA\\ 
	$^4$ Laboratory of Atomic and Solid State Physics, Department of Physics, 
	Cornell University, Ithaca, New York 14853, USA\\
	$^5$ Kavli Institute at Cornell for Nanoscale Science, Ithaca, New York 14853, USA
}
\date{\today}
 
\begin{abstract} 
 SrRuO$_3$, a ferromagnet with an approximately 160\,K Curie temperature, exhibits a $T^2$ dependent dc resistivity below $\approx$ 30 K. Nevertheless, previous optical studies in the infrared and terahertz range show non-Drude dynamics at low temperatures which seem to contradict a Fermi-liquid picture with long-lived quasiparticles. In this work, we measure the low-frequency THz range response of thin films with residual resistivity ratios, $\rho_{300K}/ \rho_{4K} \approx$ 74.   Such low disorder samples allow an unprecedented look at the effects of electron-electron interactions on the low frequency transport.  At temperatures below 30 K we found both a very sharp zero frequency mode which has a width narrower than $k_BT/\hbar$ as well as a broader zero frequency Lorentzian that has at least an order of magnitude larger scattering rate.   Both features have temperature dependencies consistent with a Fermi-liquid with the wider feature explicitly showing a T$^2$ scaling.  Such two -Drude transport sheds light on previous reports of the violation of Mathielssen's rule and extreme sensitivity to disorder in metallic ruthenates. We consider a number of possibilities for the origin of the two feature optical conductivity including multiband effects that arise from momentum conserving interband scattering and the approximate conservation of a pseudo-momentum that arises from quasi-1D Fermi surfaces.
 \end{abstract}

\maketitle

The $4d$ ruthenates are well-suited to the study of itinerant correlated electrons and the stability of the Fermi-liquid state because no explicit doping is necessary to produce metallic conduction \cite{Koster12a,Puchkov96, Maeno97,Capogna02,Perry00}.  The position of the Fermi level in bands resulting from the hybridization of O 2$p$ and Ru 4$d$ levels leads to ground-state behavior ranging from ferromagnetic order in SrRuO$_3$, metallic paramagnetism in CaRuO$_3$ \cite{Kamal06a}, antiferromagnetic insulating behavior in Ca$_2$RuO$_4$, quantum critical metamagnetism in Sr$_3$Ru$_2$O$_7$ \cite{grigera2001magnetic}, and unconventional superconductivity in Sr$_2$RuO$_4$ \cite{Perry00}. These materials present a unique opportunity to investigate a variety of correlated electron behavior in the low disorder limit.

SrRuO$_3$ exhibits a transition from a paramagnetic to a ferromagnetic state at T$_c \approx 160\,K$. Quantum oscillations and a quadratic temperature dependence of the resistivity have been measured in the highest-quality samples \cite{Capogna02, mackenzie98}. These findings suggest that the ground state of SrRuO$_3$ is a magnetic Fermi liquid. Nonetheless, among other experimental observations, infrared and optical measurements of SrRuO$_3$ films (that generally have had higher disorder levels than single crystals) have shown a finite frequency peak in $\sigma_1$ at frequencies of order 3k$_B$T \cite{Kostic98}.   At frequencies above the peak, the real part of the optical conductivity was observed to fall off as $\omega^{-1/2}$\cite{Kostic98}.    Optical measurements at lower frequency gave evidence for a related fractional power law dependence of the conductivity on the transport relaxation time \cite{Dodge00}.  The theoretical basis to understand this seeming deviation from the Lorentzian Drude form (and by implication the non-Fermi liquid nature of this material) is not clear considering the radical implications it would have on the link between ac and dc electrical transport measurements. Similar deviations from simple Drude forms of finite frequency peaks and anomalous power laws have been seen in the related compound CaRuO$_3$\cite{lee2002non,Kamal06a}. In addition, SrRuO$_3$ has a very striking negative deviation from Matthiessen's rule when impurity scattering is increased through electron irradiation. It was demonstrated that although the fractional form works for more disordered samples, it does not account for this violation for low disorder samples \cite{Klein01a,Kats02a}. These results highlight the extreme sensitivity to disorder in this material and the apparent dependence of even the inelastic scattering on sample quality.  Therefore, measuring low-disorder SrRuO$_3$ samples at low energies would shed light on the nature of electronic excitations and provide a unique opportunity to examine Fermi liquid predictions in this strongly correlated material.  Previous studies were performed at higher frequencies and/or on samples with larger disorder.

In this paper we use time domain terahertz spectroscopy (TDTS) to examine the complex conductivity and resistivity of very high quality thin films of SrRuO$_3$.  Below 30K, we find the real part of the THz conductivity exhibits two very distinct low energy peaks that are related to different conduction channels.  There is both a very sharp zero frequency conducting mode which has a width narrower than $k_BT/\hbar$ as well as a broader Lorentzian peak with at least an order of magnitude larger scattering rate.   Both features have temperature dependencies consistent with a Fermi-liquid with the wider feature explicitly showing a T$^2$ scaling.   There are a number of possibilities for the origin of these features including multiband effects that arise from momentum conserving interband scattering and the approximate conservation of a pseudo-momentum that arises from this material's quasi 1D Fermi surface sheets.

In TDTS, an approximately 1 ps long electromagnetic pulse is transmitted through a substrate and film.  The complex transmission $T(\omega) $ is obtained from the Fourier transform of the time trace, referenced to a bare substrate. Complex conductivity $\sigma(\omega) $ is calculated without the need for Kramers-Kronig transformation from the complex transmission using $T(\omega) = \frac{(1 + n)}{1+n + \sigma(\omega) d Z_0}  e^{\frac{i\omega\Delta L(n-1)}{c}}$. In this expression $n$ is the substrate index, $\Delta L$ is a correction that accounts for thickness differences between the reference substrate and the sample substrate, $d$ is the film thickness, and $Z_0$ is the impedance of free space (377 $\Omega$).  We determined the effective $\Delta L$ from a self-consistent first echo measurement of the sample and substrate at different temperatures.   The proper determination of $\Delta L$ to submicron accuracy is essential for the accuracy of these results\cite{SeeSI}.  The films were grown on single-crystal DyScO$_3$ (110) substrates by molecular-beam epitaxy (MBE) to a thickness of 23 nm\cite{nair2018synthesis}.   Because this substrate is very lossy to the THz signal in the [001] direction (see SI Fig. S3), in all presented measurements the polarization of the incident THz beam is aligned parallel to the [$\bar{1}$10] direction. Our data on samples grown on less lossy (and more highly strained) NdGaO$_3$ substrates show that the in-plane anisotropy is less than 20\% with no qualitative difference between the two directions (SI Fig. S4).


\begin{figure}
\begin{center}
\includegraphics[width=0.66\columnwidth]{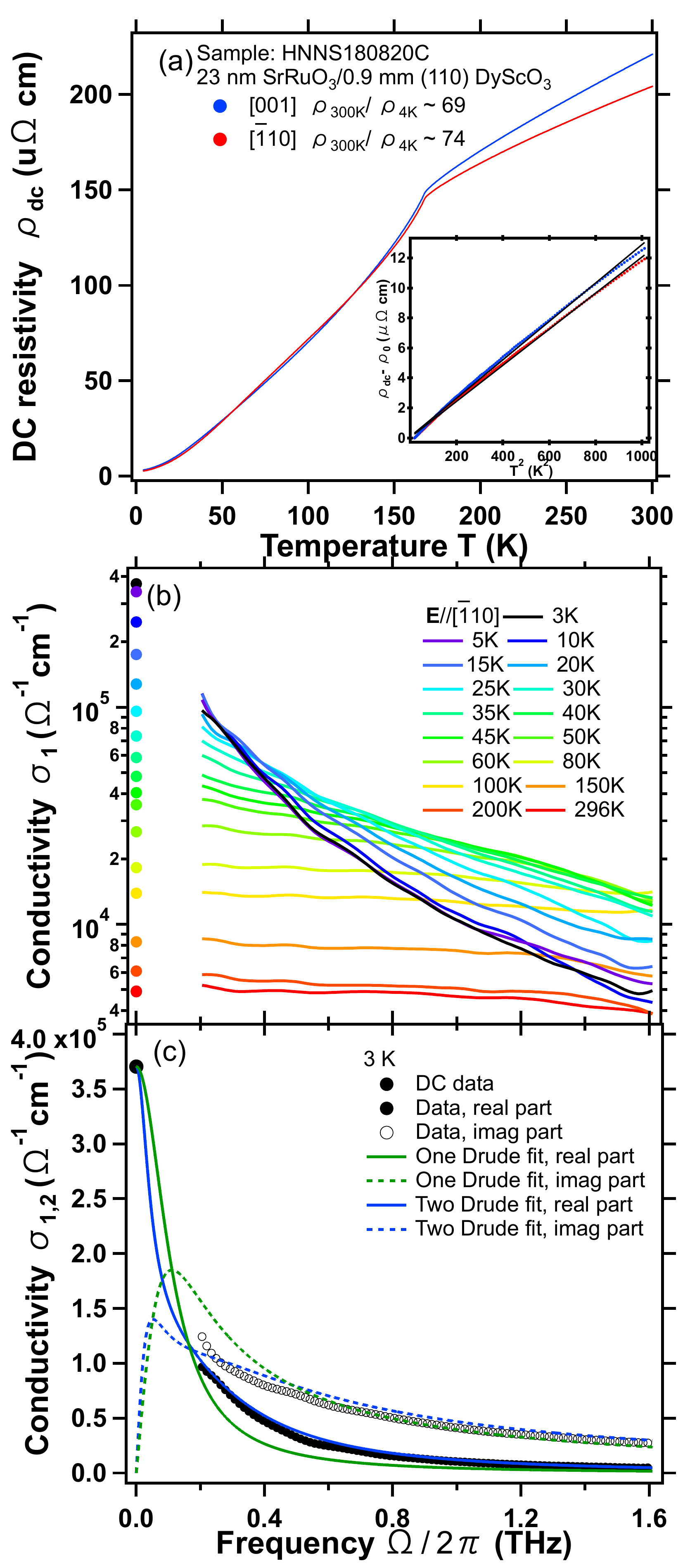}
\centering
\caption{(color online) (a) dc resistivity as a function of temperature for the SrRuO$_3$ film for the two orthogonal directions.  Inset: Resistivity minus residual resistivity as a function of T$^2$.  Fits to the data in the temperature range 2-32 K are shown as black lines. (b) Real part of the THz conductivity $\sigma_{1}$ from 5 K to room temperature along with corresponding dc values. (c) One Drude vs. two Drude fit of dc and THz data at 3 K.  The dc conductivity and real and imaginary parts of the complex conductivity are fitted simultaneously.}
\label{fig:resistivity}
\end{center}
\end{figure}

\begin{figure*}
\begin{center}
\includegraphics[width=0.8\textwidth]{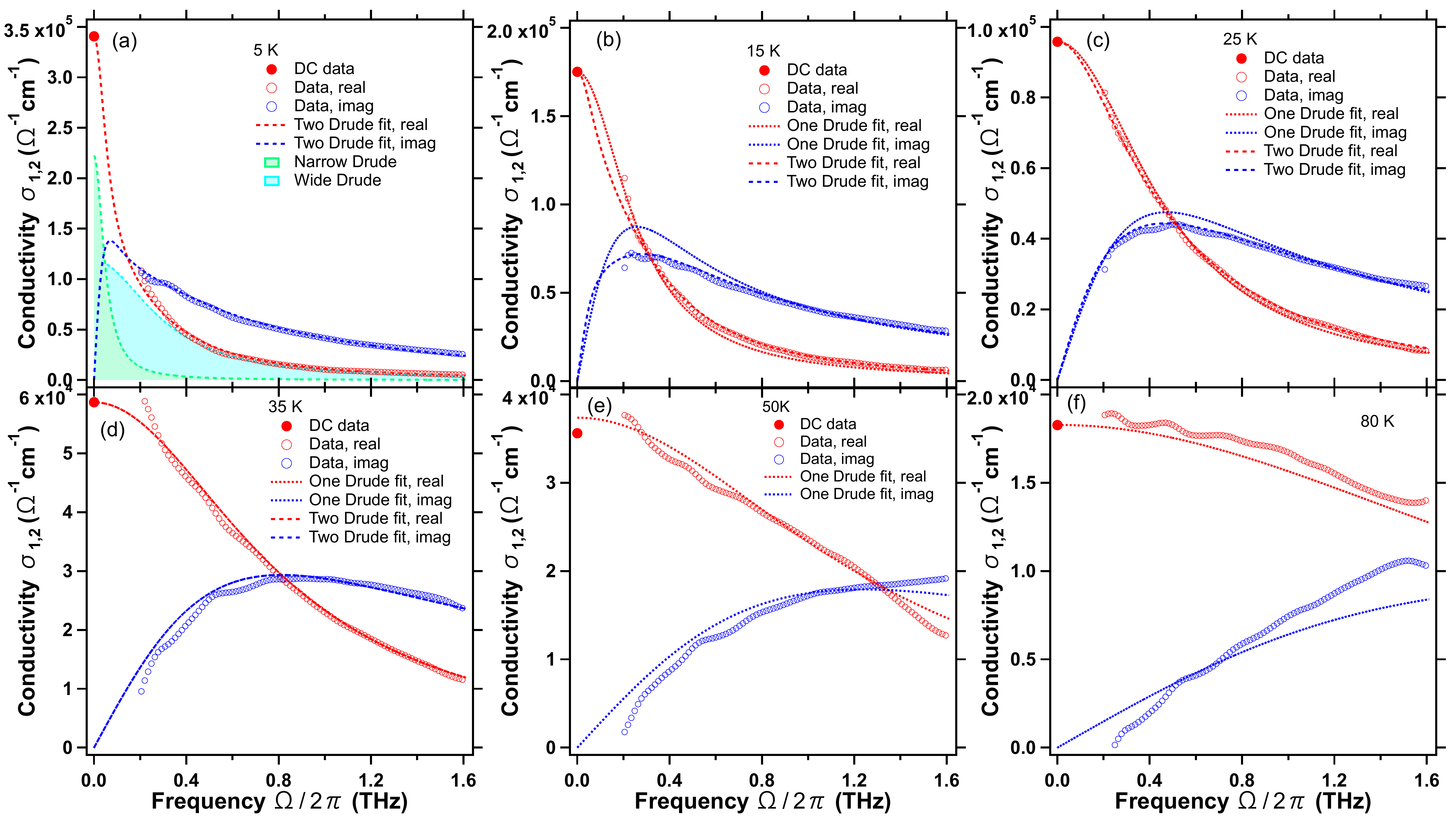}
\centering
\caption{(color online)(a) Real and imaginary THz conductivity with dc values at 5 K. The black dashed lines show modeling with two Drude terms. The green and blue shaded regions correspond to a narrow and wider Drude term, respectively. (b-d) Real and imaginary parts of the complex conductivity with dc values for 15 K and 25 K, plotted with both one and two Drude fits.  (e-f) THz and dc conductivity at (e) 35 K and (f) 80 K, plotted with a single Drude fit.}
\label{fig:srocond}
\end{center}
\end{figure*}

dc resistivity $\rho(T)$ was measured in the van der Pauw geometry and anisotropy was determined from the Montgomery method \cite{montgomery1971method} (Fig. \ref{fig:resistivity}a).  The noticeable kink at 160 K is attributed to the development of ferromagnetic order \cite{Allen96}. The high quality of the film is reflected in its low residual resistivity value of $\rho$(T$\rightarrow$0) $\sim$2.6 $\mu\Omega$ cm giving a large residual resistivity ratio of $\approx$ 74 along the [$\bar{1}$10] direction.  This residual resistivity is almost 20 times lower than the films used in previous TDTS studies \cite{Dodge00}.   A quadratic dependence on temperature of $\rho(T) -\rho(0) $ has been reported up to at least 30 K (inset of Fig. \ref{fig:resistivity}(a)).  Our observation of a T$^2$ dependence of the resistivity is consistent with other studies on low disorder samples \cite{mackenzie98, cao08} as opposed to behavior T$^\beta$, where $\beta \sim 1-2$ is observed primarily in samples with residual resistivities above 50 $\mu\Omega$ cm \cite{Allen96}. 

In Fig.  \ref{fig:resistivity}(b), we plot the real part of the THz and dc conductivity at different temperatures.   With decreasing temperature there is a remarkable sharpening of a low-frequency Drude-like peak. The real $\sigma_1$ and imaginary $\sigma_2$ parts of the complex conductivity, with corresponding dc values at 3 K are plotted in Fig. \ref{fig:resistivity}(c). At this temperature the peak in $ \sigma_1$ is so narrow that $\sigma_2 > \sigma_1$ for the entire frequency range measured.  In a simple single-band metal, one expects that the scattering of electrons is dominated by quenched disorder as T$\rightarrow0$ and that the dynamical conductivity can be modeled with a single Drude oscillator with the functional form $\sigma(\omega) = \epsilon_0  \frac{  \omega_p^2 }{1/\tau-i\omega}$ (where $\omega_p$ is the plasma frequency and $1/\tau$ is the current decay rate).  Although the $\sigma_1$ data superficially have such a Drude form, in fact the 3K complex conductivity cannot be reproduced with a single Drude oscillator.   As can be seen in Fig. \ref{fig:resistivity}(c), the best fits with a single Drude oscillator to the THz constrained with the dc resistivity underestimates the real conductivity, and overestimates the imaginary conductivity.   Note that we can make such an assessment despite the fact that the spectral range between dc and 200 GHz is not measured because the real and imaginary parts of $\sigma$ are Kramers-Kronig related to each other e.g., the data are strongly constrained for parts of the spectral range that are not explicitly measured, by parts that are measured.  In order to fit the 3K conductivity, at least two Lorentzian oscillators are needed: one narrow with  $ 1/\tau_1 \lesssim$ 50 GHz scattering rate and one wider with $ 1/\tau_2 \sim$ 300 GHz scattering rate.  Because the narrow oscillator has a width below the measured frequency range, we can only set an upper limit on its width.   In these fits, we adopt the highest value of $1/\tau_1$ consistent with THz data as its upper bound.   The full functional form is $ \sigma(\omega) = \sum_{n=1}^{2} \epsilon_0{\omega_{p_n}}^2\frac{1}{1/\tau_n-i\omega} -i\epsilon_0(\epsilon_{\infty}-1)\omega $.   Here $\epsilon_\infty$ accounts for effects of interband transitions at frequencies well above our range.

\begin{figure}
	\includegraphics[width=0.66\columnwidth]{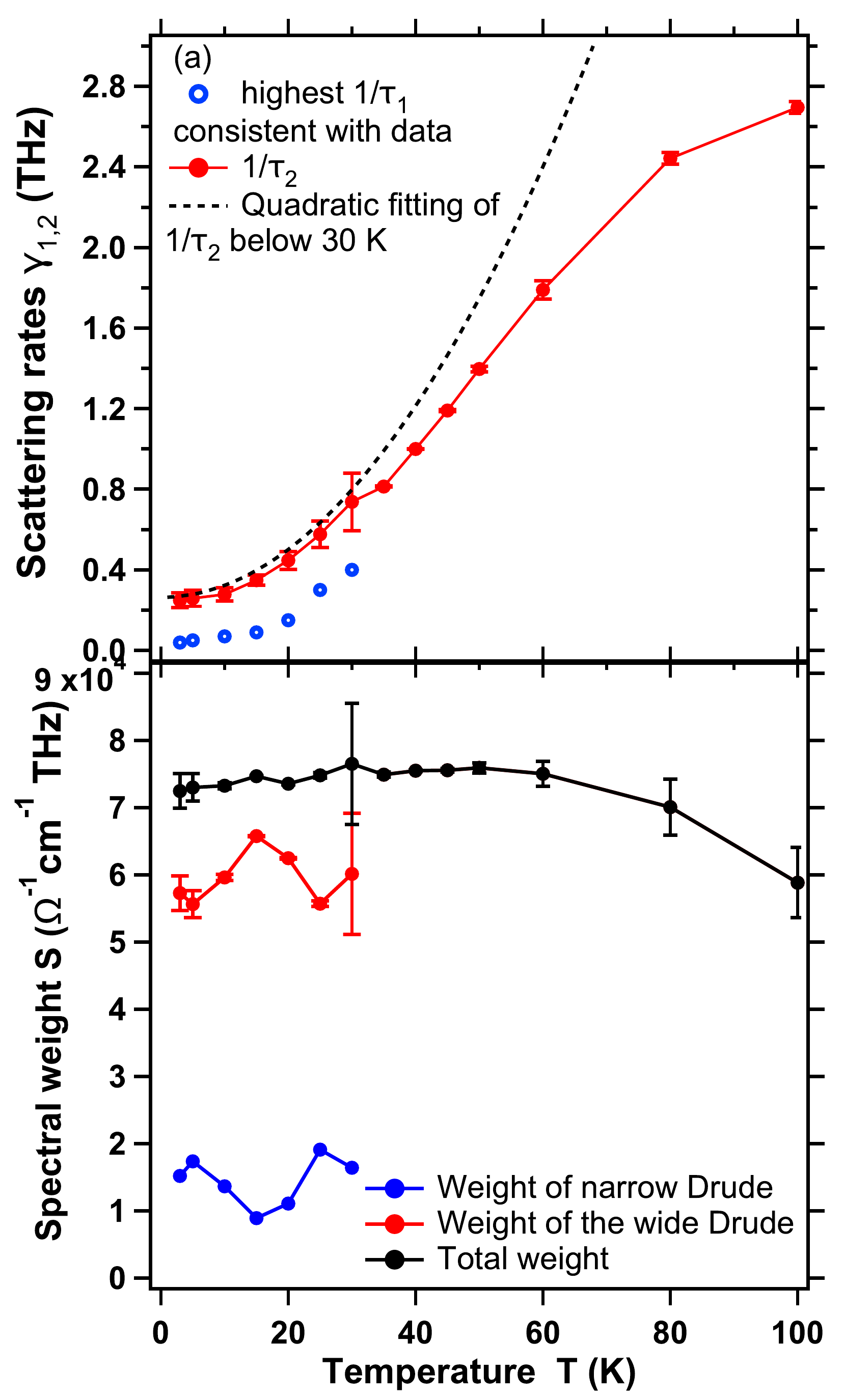}
	\centering
	\caption{(color online)(a) Scattering rate of the wide (red) and narrow (blue) Drude peaks as a function of T. The dashed line is a quadratic fit to the data below 30 K, and extended to higher T. (b) Spectral weight of the two Drude peaks (obtained from two Drude fitting) and their total spectral weight, from 5 K to 30 K.   Above 30K single Drude fit parameters are given.  The spectral weight is proportional to $\omega_p^2$.}
	\label{fig:RtWt}
\end{figure}

The necessity to use two Drude terms extends to higher temperatures.  One can see in the 5 K conductivity (Fig. \ref{fig:srocond}(a)) that the narrow Drude (green area under the real part of the narrow Drude) accounts for the sharp upturn towards the dc conductivity whereas the wide Drude part (blue) is needed to capture the long tail of $\sigma_1(\omega)$.  In Fig. \ref{fig:srocond}(b-c) in data up to 30 K, one cannot fit the imaginary part of the conductivity with a single Drude if one insists on a fit of the real part.  As the temperature increases, the scattering rate of the narrow Lorentzian increases faster than the scattering rate of the wider one and the rates become equal above 30K and the Lorentzians indistinguishable.   Hence a single Drude fitting above 30K suffices.  It is interesting to note that this temperature is close to that below which T$^2$ resistivity has been reported.  The scattering rate of the wide Drude peak goes as $T^2$ below 30 K. The total spectral weight (Fig.\ref{fig:RtWt}(b)) is unchanged within 5\% below 30 K, while the narrow Drude peak corresponds to about 20\% of the total weight in the range where it can be distinguished.   Note that the fractional functional form used previously \cite{Dodge00} does not fit our data\cite{SeeSI}.  Also note that we see no sign of meV range finite frequency peaks arising from either finite temperature effects as observed previously in higher disorder samples \cite{Kostic98}, nor that have been predicted to exist via recent density functional and dynamical mean field calculations arising from the tilt of the octahedra \cite{dang2015band}.


We examined complex resisivity which is the inverse of complex conductivity data (Supplemental Sec. V). For a single band metal, the ``Gurzhi" scaling \cite{gurzhi1959mutual} for a metal with dominant Umklapp scattering as T$\rightarrow$0 predicts the real part of the resistivity goes as $\rho_1(T,\omega) = \rho_0^{ee}(T)[1+(\hbar\omega)^2/b \pi^2 (k_BT)^2]$ where $\rho_0^{ee}(T)$ is the quadratic dc resistivity.   This relation arises from the established relation between T and $\omega$ induced inelastic scattering \cite{maslov2016optical}.   With the $b$ that is predicted to be 4 for a canonical Fermi liquid, the scattering is dominated in our temperature and frequency range by temperature e.g., the broadening at 30K at $\omega=0$ is expected to be approximately 16.6 times larger than the broadening at 1 THz at T=0.   This is consistent with our data in that we find very little frequency-dependent changes to $\rho_1$ (Supplemental FigS. 6 and 7), while the T dependence changes are large. When intraband scattering processes dominate, the complex resistivity can be related to the ``extended" Drude model (EDM), in which the scattering rate and effective masses in the Drude formula become complex and frequency dependent \cite{basov05}. The EDM has been used extensively to describe heavy fermion systems \cite{Webb86, Degiorgi99a}, high-T$_c$ cuprates \cite{Puchkov96} and transition-metal compounds \cite{Allen77}. Within the context of EDM, the slope of the imaginary part of the complex resisivtiy with frequency is proportional to the optical renormalized mass ($m^*/m_b$ ,$m_b$ is the effective band mass). It is found that in our TDTS data, mass enhancement at the lowest temperatures is $\sim$ 6.5 (Supplemental FigS. 6(b)), which roughly agrees with heat capacity and de Haas-van Alphen measurements \cite{Allen96, Alexander05}. In particular, heat capacity measurements show a Sommerfeld coefficient $\gamma_{expt}/\gamma_{theor}=3.7$, suggesting a mass enhancement similar to what we observe in TDTS measurements \cite{Allen96}.  According to an angle-resolved de Haas-van Alphen study, the effective mass of charge carriers measured for each Fermi surface sheet ranges from 4.1-6.9$\,m_e$ \cite{Alexander05}. Similarly angle-resolved photoemission has found masses of order 3.7 $m_e$ \cite{Shai13} for the $\beta$ sheet.   

We now discuss the possible origins of the narrow low frequency Drude peak. One obvious possibility is that multiple bands with varying masses near the Fermi energy contribute to two channels of conduction. Although in CaRuO$_3$ a manifold of heavy, flat bands were observed close to $E_F$\cite{liu2018revealing}, such structure has not been reported by ARPES measurements of SrRuO$_3$\cite{Shai13}, making different bands with very different velocities an unlikely source of a narrow Drude term.   Nonetheless, other possibilities exist.   Maslov and Chubukov \cite{maslov2016optical} have shown that the presence of extremely strong momentum conserving electron-electron interband scattering in the presence of weak momentum relaxing scattering
can create a double Drude structure in which the narrow Drude peak has a width which is the geometric mean of the momentum conserving and momentum relaxing scattering rates.   In a different scenario,  Rosch has shown \cite{rosch2002optical,rosch2006optical} that in 1D it is possible to define a pseudo momentum that does not decay by 2-particle collisions and hence decays more slowly than the conventional crystal momentum.   It is expected that a state with finite pseudomomentum has significant projection on current-carrying states.    This gives rise to well defined and sharp peaks in the optical conductivity.  One might expect these effects in even quasi-2D metals like SrRuO$_3$ as its Fermi surface is composed of primarily 1D sections that show only weak hybridization where the bands intersect.   Irrespective of the mechanism, what these scenarios share is the idea that scattering channels {\it add} in the conductivity not the resistivity.   The latter is the usual Mathielssen's rule and in this regard our results provide a path forward for understanding deviations from Mathielssen's rule in this material.

We have examined THz dynamical conductivity in clean films of SrRuO$_3$, which have more than an order of magnitude smaller residual resistivity that previously measured samples, and observed very different results.  At low temperature, a narrow Drude-like peak emerges which cannot be parameterized with a single oscillator.  As it is the low frequency and low temperature properties of a system which are diagnostic of its ground, the T$^2$ dependence of the widths confirm the Fermi liquid nature of this compound.  The presence of multiple Drude peaks, however, indicate effects beyond conventional Boltzmann transport and might help explain previous reports of deviations from Mathielssen's rule. They may indicate either the presence of extremely strong momentum-conserving electron-electron interactions or an almost conserved pseudo momentum due to quasi-1D Fermi surfaces of this system.

We would like to thank S. Dodge, S. Gopalakrishnan, and J. Orenstein for helpful conversations.  Work at JHU was supported though the NSF-DMR 1905519.   Research at Cornell was supported by the National Science Foundation (Platform for the Accelerated Realization, Analysis and Discovery of Interface Materials (PARADIM)) under Cooperative Agreement No. DMR-1539918. N.J.S. acknowledges support from the National Science Foundation Graduate Research Fellowship Program under Grant No. DGE-1650441. Work by D.E.S.,  J.P.R., and K.M.S. was supported by the National Science Foundation through DMR-1709255. This research is funded in part by the Gordon and Betty Moore Foundation’s EPiQS Initiative through Grant No. GBMF3850 to Cornell University. This work made use of the Cornell Center for Materials Research (CCMR) Shared Facilities, which are supported through the NSF MRSEC Program (No. DMR-1719875). Substrate preparation was performed in part at the Cornell NanoScale Facility, a member of the National Nanotechnology Coordinated Infrastructure (NNCI), which is supported by the NSF (Grant No. ECCS-1542081).  Y. Wang would like to thank Ludi Miao for argon milling of a thin strip sample for microwave measurements (not included). 

\newpage
\onecolumngrid

\setcounter{figure}{0}  
\renewcommand{\figurename}{Fig.} 
\renewcommand{\author}{}
\renewcommand{\cite}{}
\renewcommand\refname{Articles} 

\renewcommand{\title}{Supplemental Material: Sub-THz momentum drag and violation of Matthiessen's rule in an ultraclean ferromagnetic SrRuO$_3$ metallic thin film}
\begin{center}
	\textbf{\large Supplemental Material: Sub-THz momentum drag and violation of Matthiessen's rule in an ultraclean ferromagnetic SrRuO$_3$ metallic thin film}
\end{center}
\author{Youcheng Wang$^1$, G. Boss\'e$^{1,2}$, H. P. Nair$^3$, N. J. Schreiber$^3$, J. P. Ruf$^4$, B. Cheng$^1$, C. Adamo$^3$, D. E. Shai$^4$,
	Y. Lubashevsky$^1$, D. G. Schlom$^{3,5}$, K. M. Shen$^{4,5}$, and N. P. Armitage$^1$\\
	\medskip
	$^1$ The Institute for Quantum Matter, Department of Physics and Astronomy,
	The Johns Hopkins University, Baltimore, MD 21218 USA\\
	$^2$ Physics Department, University of North Florida, Jacksonville, FL 32224-7699, USA\\
	$^3$ Department of Materials Science and Engineering,
	Cornell University, Ithaca, New York 14853, USA\\ 
	$^4$ Laboratory of Atomic and Solid State Physics, Department of Physics, 
	Cornell University, Ithaca, New York 14853, USA\\
	$^5$ Kavli Institute at Cornell for Nanoscale Science, Ithaca, New York 14853, USA
}

\section{Experimental Details}
\label{Experiment}
The SrRuO$_3$ thin films studied in this work were grown on single-crystal  (110) DyScO$_3$ substrates by molecular-beam epitaxy (MBE) in a dual chamber Veeco GEN10 system to a thickness of $\sim$23 nm \cite{nair2018synthesis2}. Adsorption-controlled growth conditions are used in which an excess flux of elemental ruthenium is supplied to the growing film and thermodynamics controls its incorporation through the desorption of volatile RuOx. This growth regime minimizes ruthenium vacancies in the films and the resulting samples exhibit a high residual resistivity ratio (RRR) in transport measurements \cite{nair2018synthesis2}. The strong dependence of spectral features on sample quality highlights the necessity for such studies of utilizing oxide MBE, which produces higher quality films than those grown by pulsed laser deposition or sputtering \cite{nair2018synthesis2, Kacedon97, Chu96}.

In the technique of TDTS an infrared femtosecond laser pulse is split between two paths and excites a pair of ``Auston" switch photoconductive antennae; one acts as an emitter and the other acts as a receiver.  When the laser pulse hits the voltage-biased emitter, a broadband terahertz pulse is produced and collimated by mirrors and lenses and passes through the sample.  The terahertz pulse then falls on the receiving Auston switch.  Current only flows across the receiver switch at the instant the other short femtosecond pulse impinges on it. By varying the difference in path length between the two pulses, the entire electric field of the transmitted pulse can be mapped out as a function of time.  By dividing the Fourier transform of transmission through the sample by the Fourier transform of transmission through a reference substrate, one obtains the full complex transmission $T(\omega) $ over a frequency range that can be as broad as 100 GHz to 3.5 THz.  The complex transmission is used to calculate the complex conductivity $\sigma(\omega) $ without the need for Kramers-Kronig transformation using the expression $T(\omega) = \frac{(1 + n) }{1+n + \sigma(\omega) d Z_0} e^{\frac{i\omega\Delta L(n-1) }{c}}$.  In this expression $n$ is the index of refraction of the substrate, $\Delta L$ is a correction factor that accounts for thickness  difference between the reference substrate and the sample substrate, $d$ is the film thickness, and $Z_0$ is the impedance of free space (377 $\Omega$).  $\Delta L$ is a correction to the phase of complex transmission and thus also the complex conductivity. We determined the effective $\Delta L$ from a self-consistent measurement of the first echo of the sample and substrate at different temperatures (see Supplemental Material Section II below).


\section{$\Delta L$ determination}
\label{deltaL}

\begin{figure}
	\centering
	\includegraphics[width=0.6\textwidth]{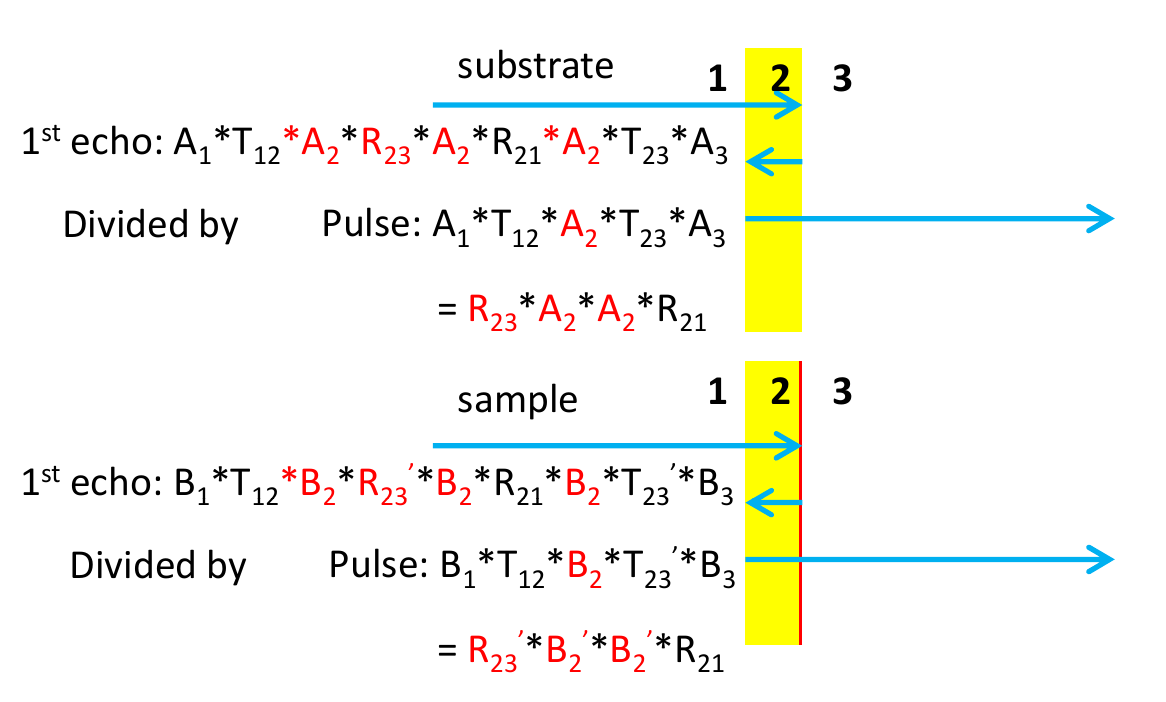}
	\caption{Phase accumulation in a first echo measurement. The blue arrows indicate the propagation of THz pulse. 1, 2, and 3 corresponds to vacuum, DyScO$_3$ and vacuum in this case. The red vertical line represents the thin film sample. Details about the symbols are explained in the text.}
	\label{fig:echo}
\end{figure}

The thickness difference between the sample substrate and a reference substrate, i.e., $\Delta L$, is usually on the order of a few microns.   Its correct determination can greatly effect the phase of calculated conductivity. A rough measurement can be made by using a micrometer on the corners of the samples, but the thickness at the center where the optical aperture is at might be different from the corners. Here we use a self-consistent measurement of the first echo of the time domain pulse to determine the value of $\Delta L$ precisely.

\begin{figure}
	\centering
	\includegraphics[width=0.95\textwidth]{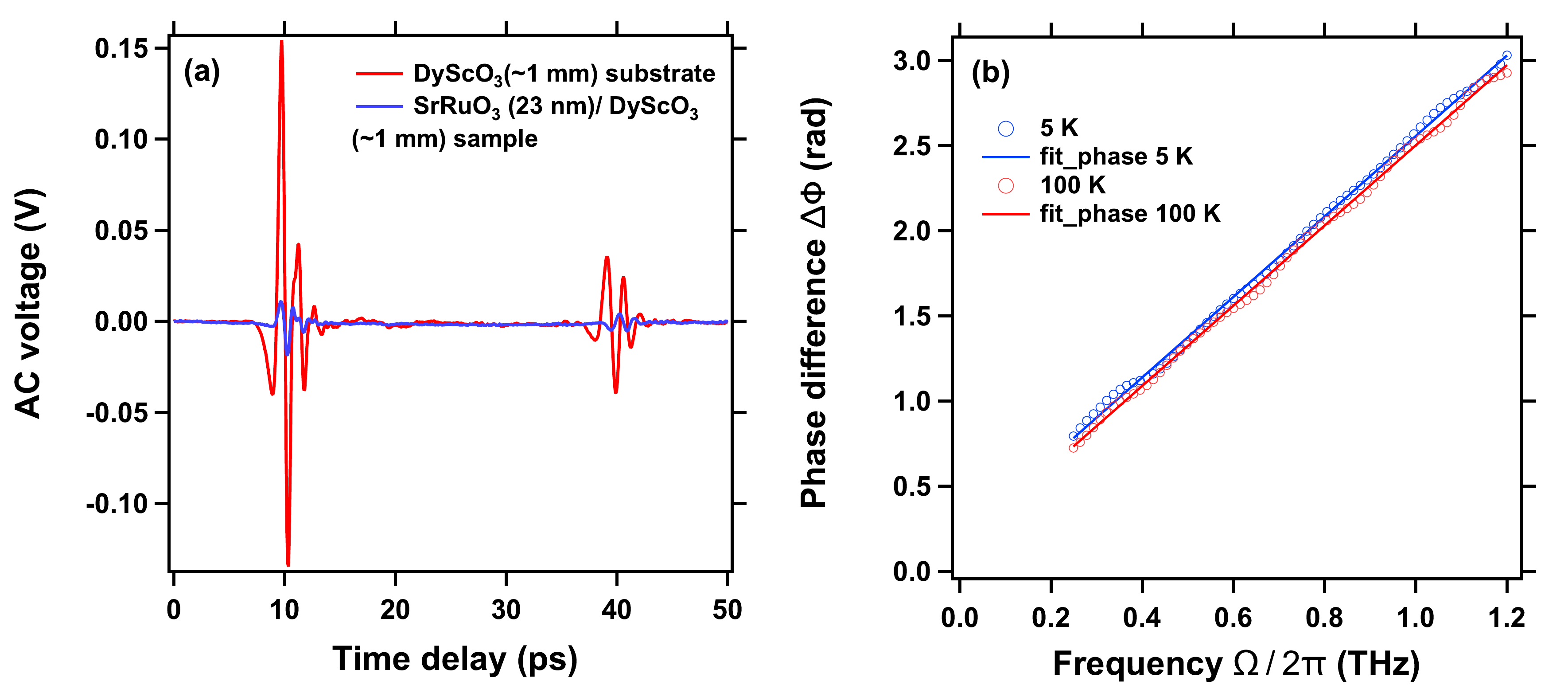}
	\caption{(a) Measured time traces of the sample (blue) and substrate (red) pulses at 5 K.  (b) Linear fits to the calculated phase difference between substrate and sample as a function of frequency. Data were taken at two temperatures and the obtained $\Delta_L$ were interpolated in between.}
	\label{fig:DeltaL}
\end{figure}

The measurement was performed by taking extended scans in time in which the transmitted THz pulse and the first echo pulse (e.g., the time delayed pulse that comes from internal reflections inside the substate) are included, for both the sample and the bare substrate. The phase accumulation can be modeled as in FigS. \ref{fig:echo}. Here $A_i$ and $B_i$ are complex phase winding in the optical path in medium $i$.  $T_{ij}$ and $R_{ij}$ are the phase shifts coefficients of transmission and reflection from the interface of medium $i$ and $j$, as determined from Fresnel equations. FigS. \ref{fig:DeltaL}(a) shows time scans of the pulses at 5 K. The transmitted and first echo pulse of the same duration ($\sim$ 20 ps) are cut from the data and then Fourier transformed, for the sample and substrate respectively. The Fourier transform of the first echo is divided by the Fourier transform of the transmitted pulse (see equations in FigS. \ref{fig:echo}). This is done for the substrate and sample respectively. The result for the sample is divided by the result of the substrate which is then, according to the equations in FigS. \ref{fig:echo}. While $R_{21}$ cancels, $R_{23}$ and $R_{23}^{'}$ are different which are given by Fresnel's equations (for example, see \cite{sushkov2010far})
\begin{equation}
\label{eq:Fresnel}
R_{23} = \frac{n_2-n_3}{n_2 + n_3} \;; 
R_{23}^{'} = \frac{n_2-n_3-y_s}{n_2 + n_3+ y_s}
\end{equation}
Here $n_2$ and $n_3$ are the indices of refraction for DyScO$_3$ (which we measure separately by referencing to an empty aperture) and vacuum, and $y_s$ is the admittance of the sample normalized with respect to 1/376.73 $\Omega$ (the admittance of vacuum). Since the admittance of the sample relies on the value of $\Delta_L$ it has to be determined through iteration until convergence (see below). The phase difference coming from the remaining factors are associated with $\Delta_L$ as in the following equation 
\begin{equation}
\label{eq:delphi}
\Delta\Phi =2 \omega \Delta L \frac{n}{c}
\end{equation}
Therefore if after subtracting off the phase factors calculated from Eq. \ref{eq:Fresnel} one fits the phase versus frequency to a straight line (see FigS. \ref{fig:DeltaL}(b)) the slope is proportional to $\Delta L$ with known or measured factors/functions including speed of light and the measured index of refraction of the substrate. Typically one starts with an initial guess coming from micrometer measurements say a few microns and calculate the conductivity and use this as input for the admittance to calculate $R_{23}^{'}$ to obtain a new value of $\Delta L$. This process is repeated for each temperature until convergence. For this sample/substrate combination $\Delta L$ converges to about 12.5 $\mu$m for both 5 K and 100 K.

\begin{figure}
	\centering
	\includegraphics[width=0.85\textwidth]{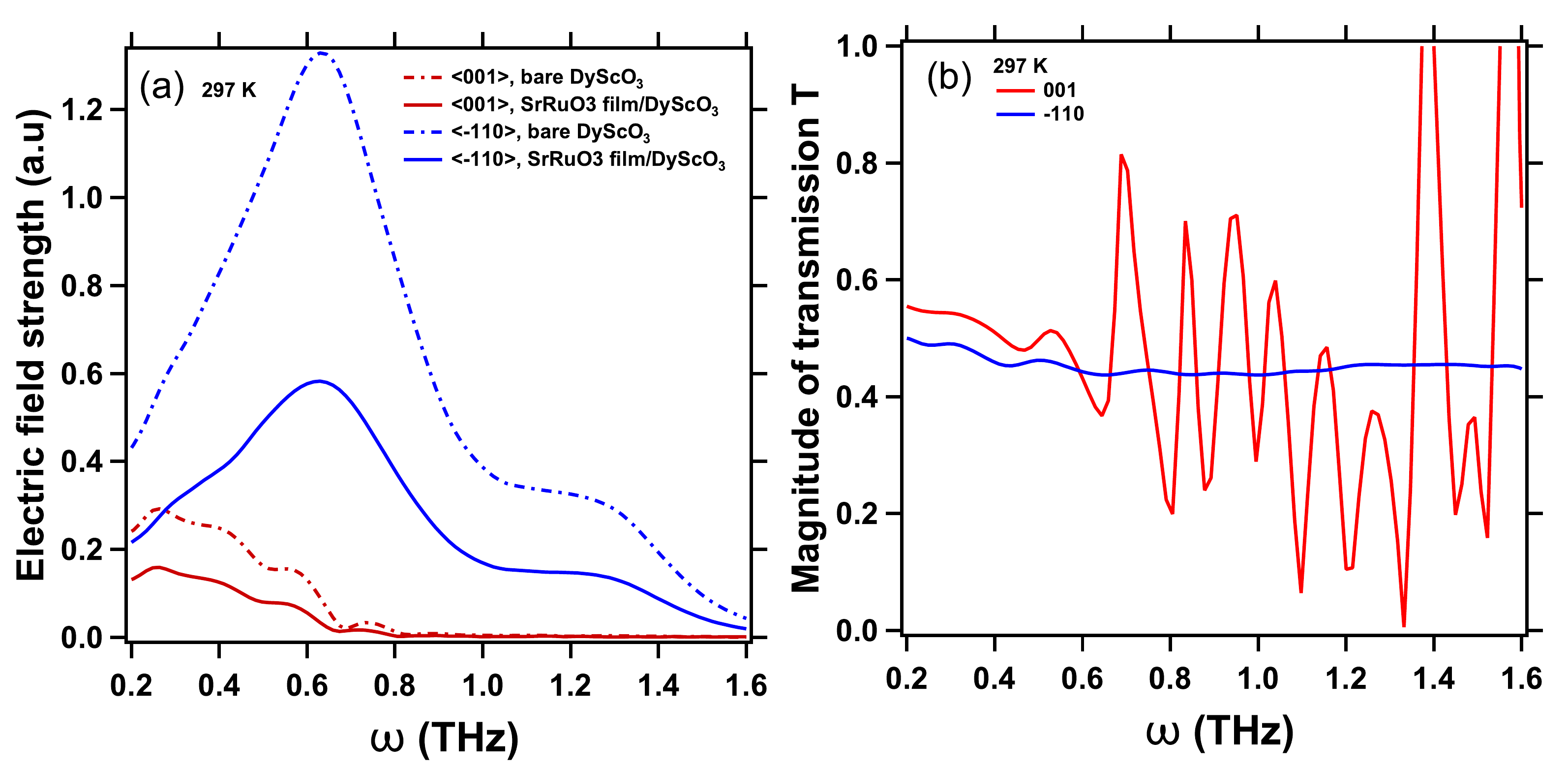}
	\caption{(a) The electric field strength as a function of frequency, through reference substrate and thin film (with substrate) along two directions. The data shown are FFTs of the time domain traces. (b) The magnitude of transmission of the sample at room temperature along two orthogonal crystallographic axes.}
	\label{fig:Ani}
\end{figure}

\begin{figure}
	\centering
	\includegraphics[width=0.47\textwidth]{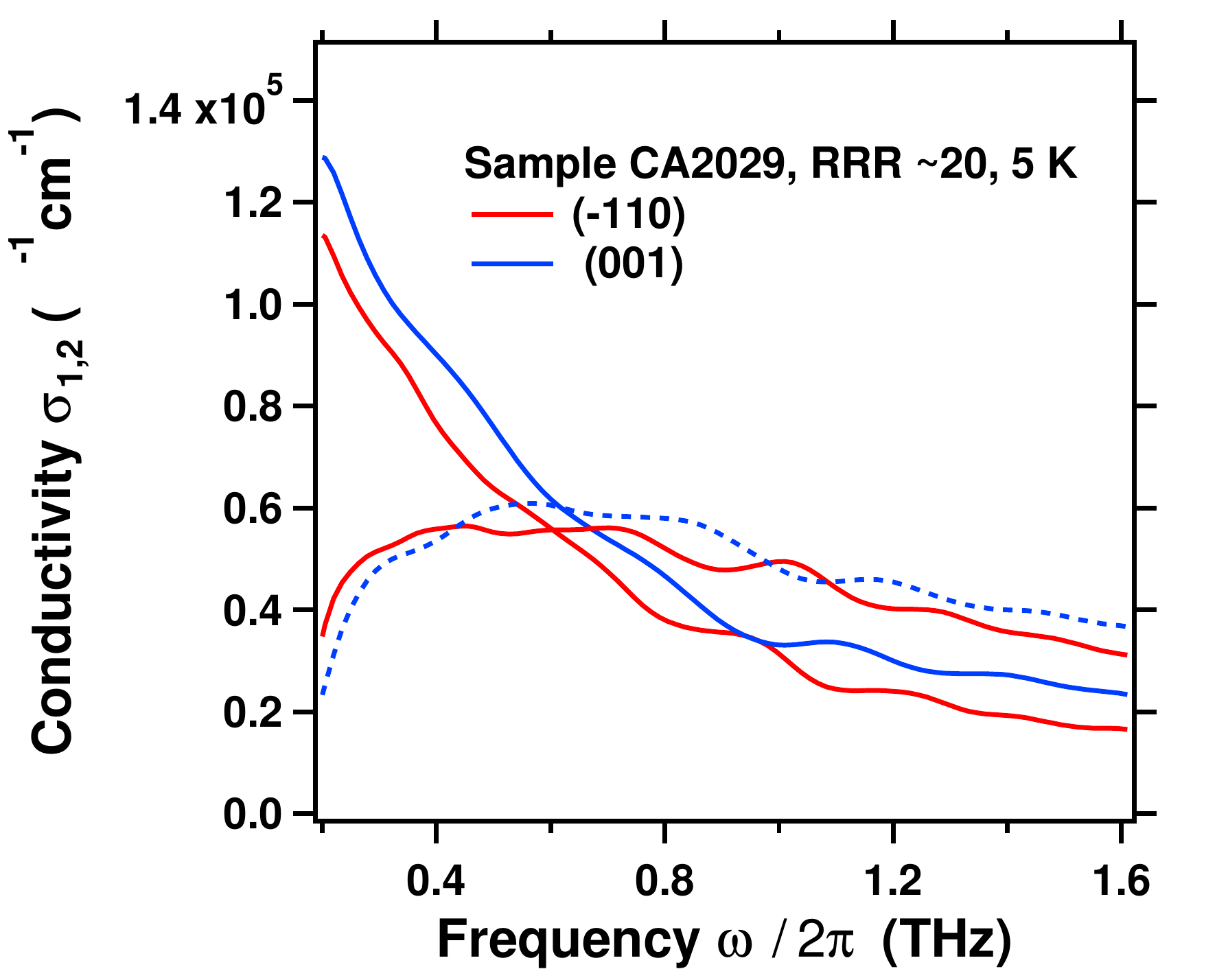}
	\caption{The room temperature conductivity of a sample grown on NdGaO$_3$ along two orthogonal crystallographic axes.  One can see the anisotropy is weak.}
	\label{fig:CA2029} 
\end{figure}
\section{Optical anisotropy}
\label{anisotropy}
As discussed in the main text, the orthorhombic DyScO$_3$ substrate has larger optical absorptions for light polarized in the $<001>$ axis than the $<-110>$ axis. See FigS. \ref{fig:Ani}(a), which shows a comparison of the magnitude of room temperature transmission $T(\omega)$ along these two orthogonal directions. One can see there is little signal above 0.8 THz for polarizations along $<001>$ axis, possibly owing to the strong absorption by polarized phonons in this frequency range.   When divides the FFT of the time trace of sample pulse by reference pulse one gets very noisy data above 0.8 THz as shown in Fig. \ref{fig:Ani}(b)).   This makes data taken with  $<001>$ polarized light unreliable for this substrate.

However, a comparison with samples grown on the NdGaO$_3$ substrate shows that anisotropy of SrRuO$_3$ is likely intrisically weak. A sample of $\sim$ 20 nm in thickness has a RRR of around 20. One can see from the 5 K data (FigS. \ref{fig:CA2029}) that there is no qualitative difference between the two directions.  Both conductivities are Drude-like in shape with slightly different scattering rates. 



\section{Fractional functional form}

\begin{figure}
	\centering
	\includegraphics[width=0.9\textwidth]{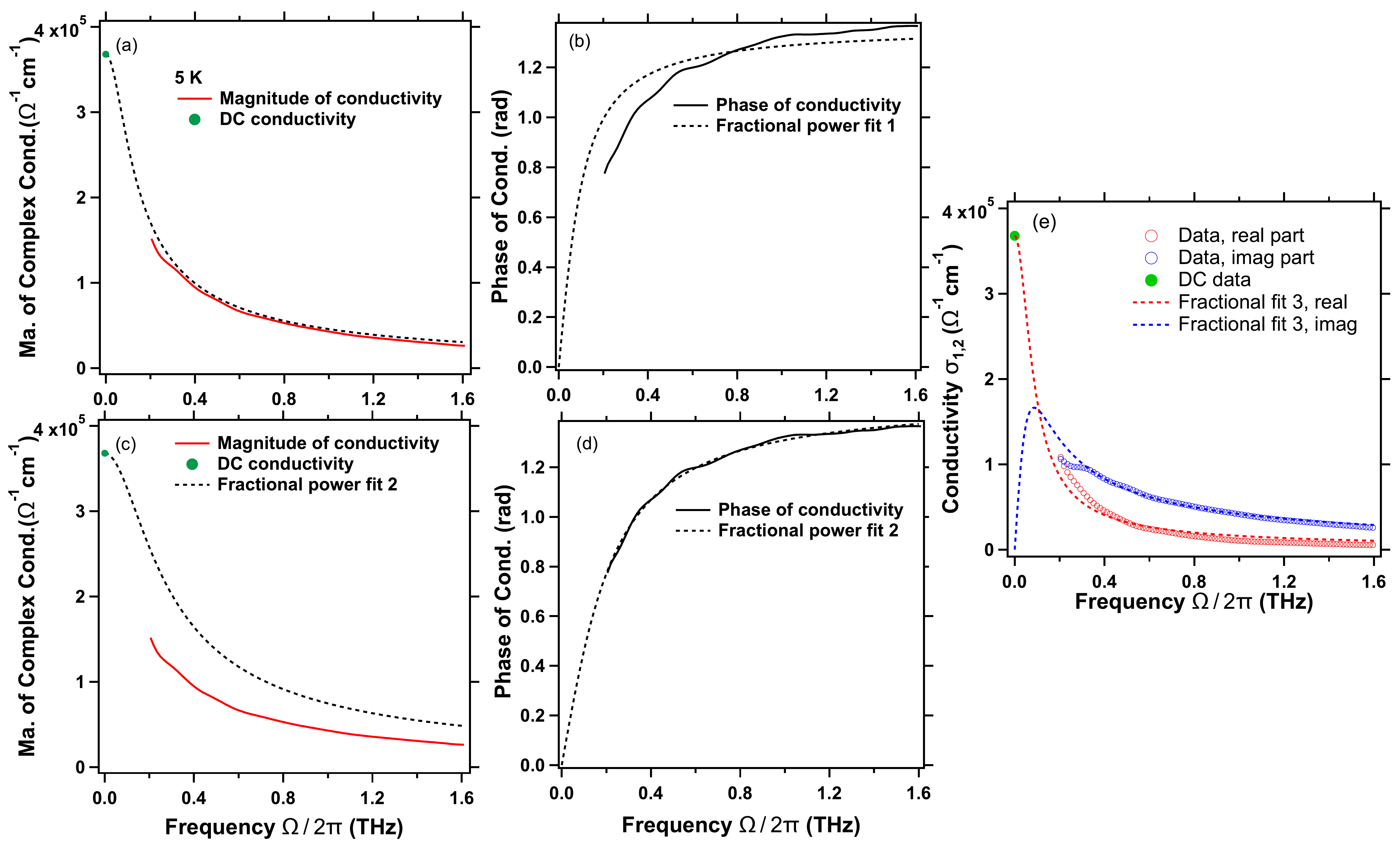}
	\caption{Fits of 5K conductivity data to Eq. \ref{FractionaForm}.  (a) and (b). A good fit to the magnitude is possible only if the fit to the phase is disregarded.  Here $1/\tau$ = 80.9 GHz and $\alpha =$ 0.853.  (c) and (d) A good fit to the magnitude is possible only if the fit to the magnitude is disregarded.  Here $1/\tau$ = 0.189 THz and $\alpha =$ 0.946. (e) A fit by converting to real and parts of the fractional power formula,  with equal weights assigned to either part. Here $1/\tau$ = 72.8 GHz and $\alpha =$ 0.803.}
	\label{fig:FracFits} 
\end{figure}

Previous work  \cite{Dodge002}  on the THz conductivity of SrRuO$_3$ used an expression of the form 
\begin{equation}
\sigma(\omega) = \frac{A}{(1/\tau - i\omega) ^\alpha}
\label{FractionaForm}
\end{equation} 
to fit the conductivity data using $\alpha =0.4$.    Although excellent fits were obtained, this was a radical proposal considering the implications it would have on the link between ac and dc electrical transport measurements.  Moreover its microscopic basis in the context of SrRuO$_3$ was not clear.  In our case, we cannot simultaneously fit both the magnitude and phase of the measured conductivity using this functional form.  This can be readily seen from FigS. \ref{fig:FracFits}.  A good fit to the magnitude is possible only if the fit to the phase is disregarded (FigS. \ref{fig:FracFits}(a-b)).   In this case $1/\tau$ = 80.9 GHz and $\alpha =$ 0.853.    A good fit to the phase is possible only if the magnitude is disregarded (FigS. \ref{fig:FracFits}(c-d)).     In this case  $1/\tau$ = 0.189 THz and $\alpha =$ 0.946. An alternative fit is done by converting the fractional formula from magnitude and phase to the real and imaginary basis, and assigning equal weight to them (FigS. \ref{fig:FracFits}(e)). In this case, $1/\tau$ = 72.8 GHz and $\alpha =$ 0.803. Note that in none of these cases is the $alpha$ exponent close to the 0.4 found in Ref. \cite{Dodge002}.   $\alpha = 1$ is the Drude limit.

\label{Fractional}


\section{Complex resistivity and the extended Drude analysis}

\begin{figure}
	\includegraphics[width=0.40\columnwidth]{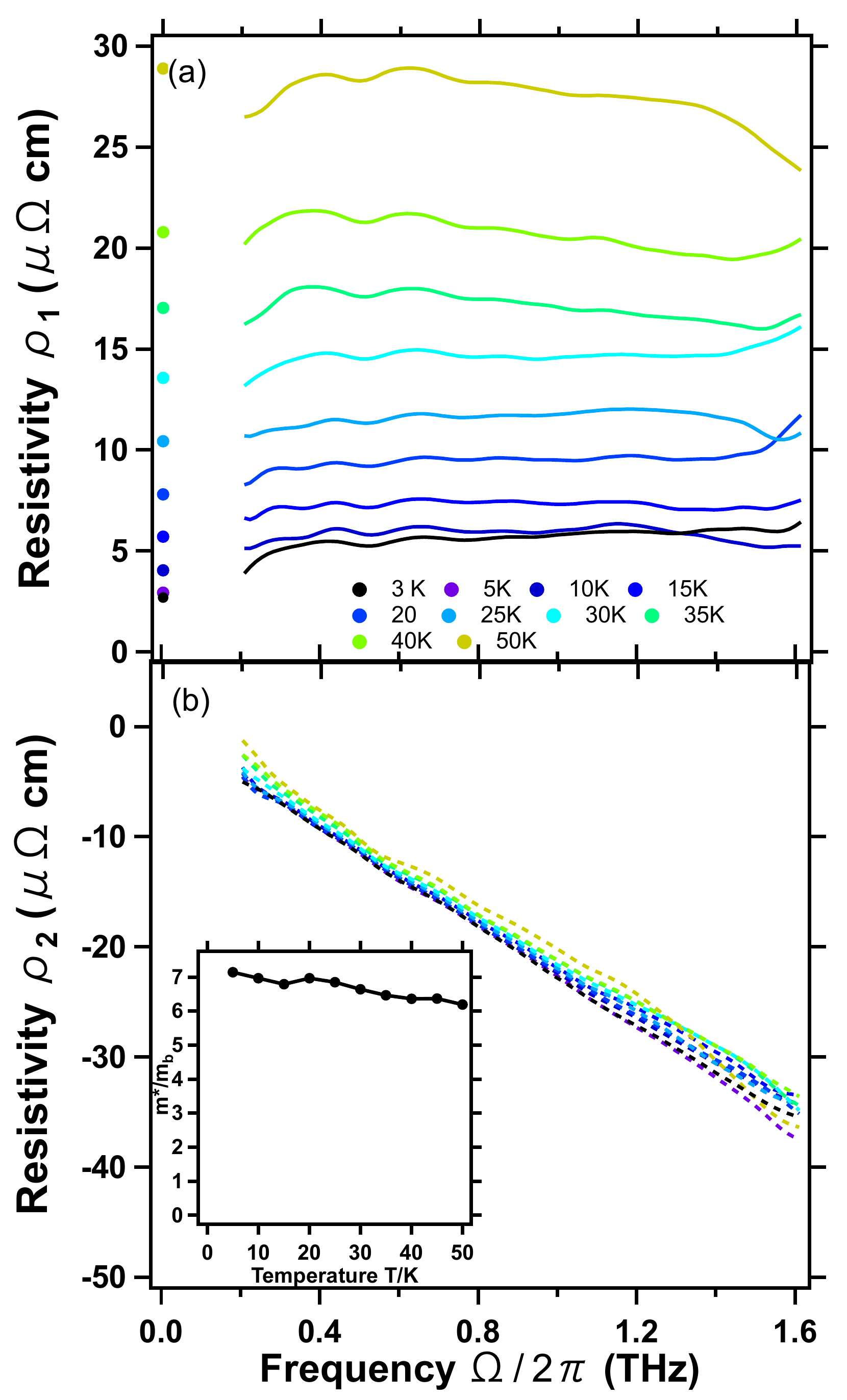}
	\centering
	\caption{(color online) (a,b) Real and imaginary parts of the complex THz resistivity at different temperatures. The real part of the resistivity is related to scattering rate while the slope of the imaginary part is a measure of the mass renormalization. The inset of (b) shows the frequency-averaged $m^*$ as a function of temperature up to 50 K.}
	\label{fig:Res}
\end{figure}
\begin{figure}
	\centering
	\includegraphics[width=0.45\textwidth]{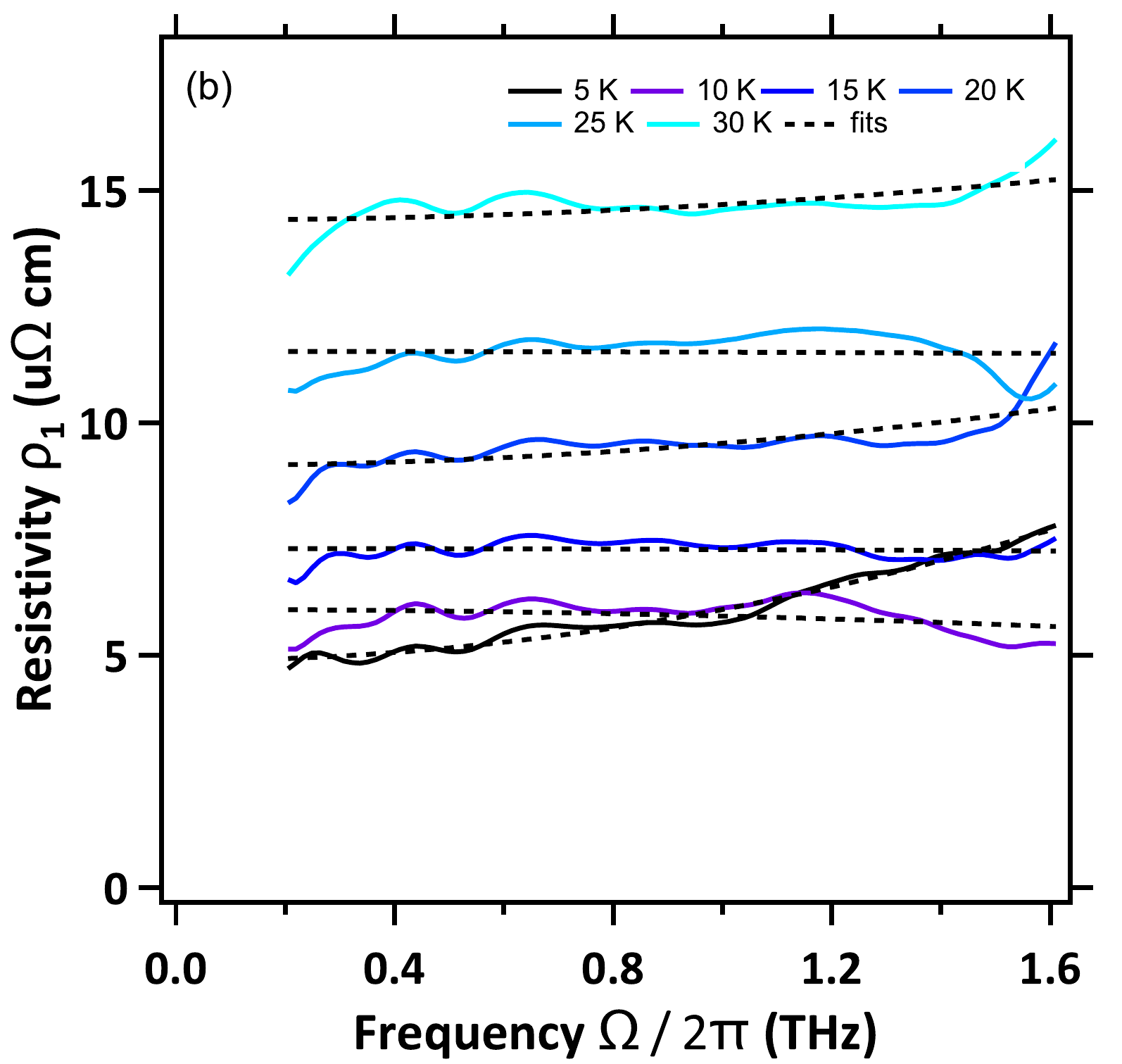}
	\caption{Real part of the complex resistivity$\rho_1(\omega)$ and tentative quadratic fitting.}
	\label{fig:Rho1fit}	
\end{figure}

FigS. \ref{fig:Res} shows real and imaginary parts of complex optical resistivity calculated from inverting the conductivity. The real part of complex resistivity is plotted along with dc resistivity.  The real part of resistivity $\rho1$ at THz frequencies has only a weak frequency dependence, which is consistent with the scale of the relative contributions of temperature and frequency dependent scattering from the Gurzhi scaling \cite{gurzhi1959mutual2}. At the lowest measured frequencies and low temperatures, there is a small drop in resistivity to the dc values, owing to the narrower Drude term.  Nevertheless, we can still try to set bounds on the size of the $\omega^2$ dependence to obtain an upper bound for the value of coefficient $A$ from the general scaling relation $\rho_1(T,\omega) = \rho_0^{ee}(T)[1+(\hbar\omega)^2/b \pi^2 (k_BT)^2]$ where $\rho_0^{ee}(T)$  mentioned in the main text. In FigS. \ref{fig:Rho1fit} we show the close up view of $\rho_1$ as plotted in Fig. 4(a) of the manuscript.  The quadratic fits are presented as dashed black lines for different temperatures. Of the positive values, b = 2.3 for 5 K,b =5.5 for 20 K, and b = 7.7 for 30 K, respectively. This is a size roughly consistent with Gurzhi scaling where b  =4 (the assumption that two particle Umklapp scattering dominates transport).  

We performed extended Drude analysis, specifically to estimate mass enhancement. The result is shown in the inset of FigS. \ref{fig:Res} (b).
Rewriting the complex conductivity $\sigma(\omega)$ in terms of a complex memory function, one obtains 
\begin{equation}
\label{eq:edmcond}
\sigma(\omega) =\epsilon_0 \omega_p^2\frac{1}{1/\tau(\omega) -i\omega[1+\lambda(\omega)]}
\end{equation}
\noindent where, adopting Boltzmann style terminology, 1/$\tau(\omega) $ and $1+\lambda(\omega) $ describe a frequency dependent scattering rate and mass enhancement ($m^*/m_b$) of the optical excitations due to many-body interactions \cite{Puchkov962}.  Here $\omega_p /2\pi = 25,000$ cm$^{-1}$ is the plasma frequency from Ref. \cite{Kostic982} and $m_b$ is the effective band mass \cite{Comment1}.  Equivalently, the imaginary part of the resistivity $\rho_2$ (FigS. \ref{fig:Res}(b)), can be used to estimate the magnitude of the effective mass by taking the slope of $\rho_2$ multiplied by $-\epsilon_0 \omega_p^2$ (inset of FigS. \ref{fig:Res}(b)). 

'
\label{rhofitting}

\section{Attempts of Microwave spectroscopy Measurements}
In the technique of Corbino broadband microwave spectroscopy \cite{liu2013broadband}, a coax cable connected to a network analyzer is terminated by a thin film sample (via a customized adapter that allows the sample to be press fit onto the end of the coax).  Complex reflection coefficients are measured at each frequency of a scan, from which impedance can be calculated from the matching equations. A linear calibration scheme is used to obtain the actual reflection coefficient from the sample alone. In the thin film limit impedance can be inverted to obtain conductance.    This technique is most sensitive when measuring samples that have impedances close to 50 $\Omega$.

We attempted to measure thin films of SrRuO$_3$ in this technique's 10s of MHz to approximately 10 GHz range in order to partially fill in the missing frequency range between dc and the low end of the TDTS experiments.   However these experiments proved to be very challenging.  The issue is that the films are so highly conductive (impedances near 0.4 $\Omega$) that their reflection coefficients are close to -1 making them hard to distinguish from a perfect conductor.  In order to enhance the impedance of the highly conductive sample so that the reflection coefficient is farther from -1, the film is patterned into a thin strip geometry \cite{scheffler2007strip}. First a Au (200 nm)/Ti (5 nm) contact was evaporated on the sample using a macroscopic donut shaped (or Corbino) mask. Then the entire film was spin coated with about 2 microns of positive photo-resist and baked at 120$^\circ$ for 1 min to harden it. Afterwards, a thin strip mask, which was actually a thin Copper wire was attached to a glass slide (could be a micro-structure on photo mask), was aligned and pressed against the center of the donut shaped Au/Ti structure, and exposed to UV. The film was developed and rinsed with DI water, and re-baked for 1 min at 120$^\circ$, so that a a thin strip (~70 microns) of photo-resist was left on top of the sample. This strip made of photo-resist is then used as as a mask for Ar milling. The impedance was enhanced from this thin strip patterning by a factor of 22.   Through this procedure, we attempted to measure the conductivity of thin strip samples.

Unfortunately, we believe that these efforts to measure the conductivity in this range largely failed.     We always found a large and unexplained mismatch between THz and microwave data. Higher temperature ($>$ 30 K) data was used as a substrate correction \cite{liu2013broadband}, assuming that the conductivity has a scattering rate much greater than 10 GHz according to extrapolation from THz data. However, it was found that the low temperature conductivity (especially $<$ 20 K) exhibited a narrow Drude-like peak feature only of the order of a few GHz wide, which is much less than simple extrapolations of the THz to the dc conductivity. This result would imply a finite frequency peak in conductivity in the 50 GHz-150 GHz range.   Although in principle this is possible, this would violate any simple picture of extrapolating the THz data to the dc data and it is hard to envision any scenario in which this occurs.   Moreover, such a scenario would give total spectral that that was strongly nonconserved as a function of temperature.  Therefore we believe that this finite frequency peak is an artifact of still having a very low impedance and therefore considerable experimental uncertainty.  Therefore, we could not reliably obtain the microwave conductivity, partly because the film is highly conductive and that contact impedance is difficult to model. 


\begin{thebibliography}{13}
\bibitem[{Com()}]{Comment1}
	\bibinfo{note}{Note that this is a slightly smaller value than the 29100
		cm$^{-1}$ predicted from band structure calculations
		\cite{Mazin97a2,Mazin06a2}.}
\end{thebibliography}

\end{document}